\documentclass[aps,prb,preprint,floatfix,superscriptaddress,showpacs,amssymb,amsmath]{revtex4}

\usepackage{graphicx}

\begin{document}

\title{Voltage-tunable ferromagnetism in semimagnetic quantum dots with few
particles: magnetic polarons and  electrical capacitance}
\author{Alexander O. Govorov}
\affiliation{Department of Physics and Astronomy, Condensed Matter
and Surface Science Program, \\ Ohio University, Athens, Ohio
45701-2979, USA}

\date{\today } 

\begin{abstract}
Magnetic semiconductor quantum dots with a few carriers represent
an interesting model system where ferromagnetic interactions can
be tuned by voltage. By designing the geometry of a doped quantum
dot, one can tailor the anisotropic quantum states of magnetic
polarons. The strong anisotropy of magnetic polaron states in
disk-like quantum dots with holes comes from the spin splitting in
the valence band. The binding energy and spontaneous magnetization
of quantum dots oscillate with the number of particles and reflect
the shell structure. Due to the Coulomb interaction, the maximum
binding energy and spin polarization of magnetic polarons occur in
the regime of Hund's rule when the total spin of holes in a
quantum dot is maximum. With increasing number of particles in a
quantum dot and for certain orbital configurations, the
ferromagnetic state becomes especially stable or may have broken
symmetry. In quantum dots with a strong ferromagnetic interaction,
the ground state can undergo a transition from a magnetic to a
nonmagnetic state with increasing temperature or decreasing
exchange interaction. The characteristic temperature and
fluctuations of magnetic polarons depend on the binding energy and
degeneracy of the shell. The capacitance spectra of magnetic
quantum dots with few particles reveal the formation of polaron
states.
\end{abstract}


 \maketitle

\section{Introduction}

Diluted magnetic semiconductors combine high-quality crystal
structures with the magnetic properties of impurities
\cite{Furdyna} and represent an important class of materials for
spintronics and quantum information \cite{General}. The
ferromagnetic ordering in diluted magnetic semiconductors can come
from the carrier-mediated interaction between magnetic ions
\cite{Zener,VFerro,M,D}. Since the carrier density in
semiconductor field-effect transistors is a voltage-tunable
parameter, the ferromagnetic state of the impurities coupled to
the carriers also becomes controlled by the voltage \cite{D}.
Voltage-control of the ferromagnetic phase transition has been
already demonstrated for the $Si/Ge$ and $A_3B_5$ material systems
\cite{VFerro}. This ability to externally control the properties
of magnetic crystals with means other than the external magnetic
field may have important device applications.

An important feature of modern nanotechnology is the ability to
shape semiconductor crystals, designing their quantum properties.
Quantum confinement of carriers is expected to strongly affect the
magnetic properties of crystals since the quantum-confined
structures can be designed in a way to strongly localize carriers
near magnetic impurities. The first step toward quantum
confinement has been made with the quasi-two-dimensional
structures where unusual ferromagnetic properties have been
described \cite{D,Ferro2D1,Ferro2D2,Dima}. For one-dimensional
lithographic structures, it was found that their transport
properties are controlled by the domain walls \cite{Molenkamp}. In
parallel, self-organization growth technology suggests
zero-dimensional nano-size quantum dots (QDs) \cite{QD1,QD2} which
can locally store carriers. Moreover, it has been demonstrated in
many experiments
\cite{QDVoltage1,QDVoltage2,QDVoltage3,QDVoltage4} that the
numbers of carriers and wave function of a QD can be changed by a
voltage applied to specially-designed metal contacts; this may
permit manipulation of ferromagnetic states in QDs by voltage.
Therefore, the combination of semiconductor QDs with magnetic
impurities looks particularly interesting; information in such
magnetic QDs can be stored not only in the number of carriers but
also in the form of the Mn magnetization. Currently, magnetic QDs
are a hot topic
\cite{Bhattacharjee0,MagDotsSingleMn,Bhattacharjee,Brey,Govorov1,Govorov2,Peeters,Pawel2004,Kulakovskii-PRB,Cincinnati}.
One important property of Mn-doped nanostructures is that a single
particle (electron or hole) can strongly alter the ground state of
the system, leading to formation of a magnetic polaron
(MP)\cite{Dima,Kulakovskii-PRB,Cincinnati}. In the case of a QD
with a single Mn impurity, single carriers lead to the formation
of hybrid electron-Mn states
\cite{Bhattacharjee,Govorov1,Govorov2,Pawel2004}.

In a semi-magnetic QD, a localized MP is formed due to the
exchange interaction between  the spins of Mn ions and a carrier
trapped in a QD. The MP localized inside a QD resembles a
localized acceptor-bound exciton in a bulk semiconductor doped by
magnetic impurities \cite{MPBulk,MPBulk2}. However, self-assembled
QDs have important differences: (1) a single QD can trap several
electrons (holes), (2) the number of particles in a QD can be
tuned with the voltage applied to a metallic contact, and (3) the
confining potential of QDs typically is very different from the
Coulomb potential.

Here we describe anisotropic MP states of doped QDs of cylindrical
symmetry with few holes. The binding energy and magnetization of
MP states demonstrate oscillations as a function of the number of
holes due to the shell structure. The maximum binding energy and
magnetization occur in the regime of the maximum total spin of the
hole subsystem when Hund's rule is applied. Note that a recent
paper \cite{Brey} describes only the odd-even parity oscillations
in magnetic QDs (these oscillations come from the spin
susceptibility of a system with a discrete spectrum). The enhanced
Mn polarization and strong MP binding in a QD with cylindrical
symmetry described here originate from the symmetry of the system
and the Coulomb interaction. For QDs with few particles we also
predict transitions from magnetic to non-magnetic states when the
temperature or exchange interaction varies. In addition, we focus
here on the hole-mediated ferromagnetism in disk-shaped QDs where
the MP state is strongly anisotropic: the spontaneous
magnetization appears preferentially in the growth direction.

Another interesting question related to nano-scale ferromagnetism
is how the coupled Mn-hole system develops from the MP behavior
toward the Zener ferromagnetic phase transition regime in the
limit of a large number of particles. To address this question, we
will also consider spin fluctuations in a QD as a function of the
hole number and a formal self-consistent solution with critical
behavior. These results may help to answer the above question. The
crossover from the MP regime toward the phase transition behavior
can also be important from the point of view of device
applications. For example, it is essential to estimate the minimal
number of carries needed to achieve a stable ferromagnetic state
in a single QD.

The paper is organized as follows: Section II presents a model of
a self-assembled QD, Sections III and IV describe the anisotropic
MP state with one hole, Section V, VI, and VII contain the results
on few-hole QDs, and Sections VIII, IX, and X discuss critical
phenomena, fluctuations, and electrical capacitance of magnetic
QDs.

\section{Model}

We model the hole-Mn complex in a self-assembled diluted
semiconductor QD with the following Hamiltonian:

\begin{eqnarray}
\hat{H}_{hh}=\frac{\hat{{\bf p}}^2}{2 m_{hh}}+U({\bf R})-
\frac{\beta}{3}\hat{j}_{z}\hat{S}_z, \label{Hamiltonian}
\end{eqnarray}
where ${\bf R}=({\bf r},z)$ is the radius vector, $z$ is the
vertical coordinate, ${\bf r}=(x,y)$, ${\bf p}$ is the momentum,
 and $\hat{j}_{z}$ and $\hat{S}_z$
are the z-components of the hole and Mn momentums, respectively.
$\hat{S}_z=\sum_i\hat{S}_{i,z}\delta({\bf R}-{\bf R}_i)$, where
$\hat{S}_{i,z}$ and ${\bf R}_i$ are the spin and position of the
$i$-impurity, respectively. It is convenient to model the in-plane
motion of a hole by the parabolic potential and the vertical
motion with a square well \cite{book}. So, we write: $U({\bf
R})=u(z)+m\omega_0^2r^2/2$, where $u(z)$ is the z-confinement
potential and $\omega_0$ is the in-plane frequency. The
anisotropic exchange interaction in eq.~\ref{Hamiltonian} implies
that the QD is disk shaped and the vertical size of QD, $L$, is
much smaller than the in-plane wave function dimension,
$l=\sqrt{\hbar/m_{hh}\omega_0}$, i.e. $L\ll l$. Since we consider
the only heavy-hole states, the exchange interaction (the last
term in eq.~\ref{Hamiltonian}) becomes strongly anisotropic
\cite{Mn-electron}. In our model, the light-hole states are
assumed to be strongly split from the lowest heavy-hole states in
the QD. The single-particle spatial hole wave functions and their
energies in the absence of Mn impurities are given by
$\psi_{n,m}=f_0(z)\chi_{n_x,n_y}({\bf r})$ and
$\epsilon_{n,m}=\hbar\omega_0(n_x+n_y+1)$, respectively. Here, the
wave function $f_0(z)$ corresponds to the lowest state in a square
potential well in the z-direction, and $\chi_{n_x,n_y}({\bf r})$
are the usual wave functions of a 2D harmonic oscillator;
$n_{x(y)}=0,1,2,...$ are the quantum numbers.

\begin{figure}[tbp]
\includegraphics*[width=0.6\linewidth,angle=90]{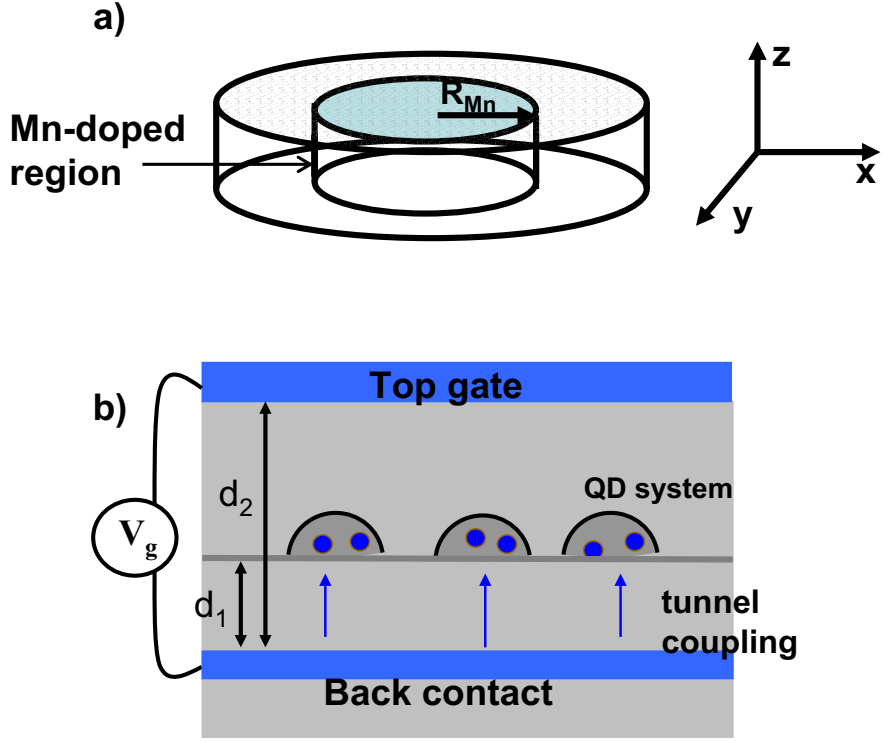}
\caption{(a) Model of a Mn-doped self-assembled QD. (b) Transistor
structure with magnetic QDs; the gate voltage controls the number
of holes in the QD layer  and therefore the magnetic state of QDs.
Capacitance of this structure can reveal the magnetic state of
QDs. } \label{fig1}
\end{figure}

In the spirit of the mean field theory, we can average the
operator (\ref{Hamiltonian}) over the impurity positions and
write:

\begin{eqnarray}
\hat{H}_{hh}'=\frac{\hat{{\bf p}}^2}{2 m_{hh}}+U({\bf
R})-\frac{\beta}{3} x_{Mn}({\bf R})N_0\hat{j}_{z}\bar{S}_z({\bf
R}), \label{Hamiltonian1}
\end{eqnarray}
where $x_{Mn}({\bf r})$ is the reduced Mn spatial density in the
system, $N_0$ is the number of cations per unit volume, and
$\bar{S}_z({\bf R})$ is the locally averaged Mn spin:

\begin{eqnarray}
\bar{S}_z({\bf R})=S B_S(\frac{\beta/3\bar{j}_z({\bf
R})}{k_B(T+T_0)}), \label{Brillion}
\end{eqnarray}
where $S=5/2$, $B_S$ is the Brillouin function, $\bar{j}_z({\bf
R})=<\Psi(R',\chi)|\hat{j}_z\delta(R-R')|\Psi(R',\chi)>$ is the
averaged momentum of the hole at the Mn position, and
$|\Psi(R,\chi)>$ is the wave function of a hole which depends on
the spatial and spin coordinates, $R$ and $\chi$, respectively;
$\chi=\pm3/2$; $T_0$ in eq.~\ref{Brillion} appears due to the
anti-ferromagnetic interaction between Mn ions.

\section{Mean field solution}

According to eq.~\ref{Hamiltonian1}, the hole in a magnetic QD
moves in the presence of the effective spin-dependent potential:

\begin{eqnarray}
U_{eff}=U({\bf R})-\frac{\beta}{3} x_{Mn}({\bf
R})N_0\hat{j}_{z}\bar{S}_z({\bf R}). \label{Ueff}
\end{eqnarray}
In a QD with strong spatial confinement, we can neglect the effect
of the second term in eq.~\ref{Ueff} on the spatial wave function.
At the same time, we should keep it for the spin part of the wave
function. The ground state of a magnetic QD with a single hole has
a simple form:

\begin{eqnarray}
\Psi=\psi_{0,0}|\uparrow>, \hskip0.3 cm \bar{S}_z({\bf R})=S
B_S(\frac{\beta/3\bar{j}_z({\bf R})}{k_B(T+T_0)}), \hskip0.3 cm
\bar{j}_z({\bf R})=\frac{3}{2}\psi_{0,0}(R)^2, \label{mpolaron1}
\end{eqnarray}
where $|\uparrow>$ is the hole state with $j_z=+3/2$. Since
$\beta<0$ for the Mn-hole interaction, $\bar{S}_z({\bf R})<0$: at
low $T$, the spins of the hole and Mn ions inside the QD are
anti-parallel. This is a state of magnetic polaron with the
energy: $E_0(T)=\hbar\omega_{0}-E_{b}(T)$. The second term in the
above equation plays the role of the MP binding energy:

\begin{eqnarray}
E_{b}(T)=-\frac{\beta}{3}\int_R{d^3R [\psi_{0,0}^2 x_{Mn}({\bf
R})N_0\frac{3}{2}SB_S({\bf R},T)]}. \label{MPolaronBEnergy}
\end{eqnarray}
In our definition, the binding energy $E_{b}(T)<0$. The ground
state of a MP is two fold degenerate since the states $j_z=\pm3/2$
have the same binding energy.

The total Mn polarization is calculated as

\begin{eqnarray}
S_{tot}(T)=\int_Rd^3R [x_{Mn}N_0 S B_S(\frac{\beta/3\bar{j}_z({\bf
R})}{k_B(T+T_0)})]. \label{IMntot}
\end{eqnarray}
The corrections to the wave function (\ref{mpolaron1}) and the
energy (\ref{MPolaronBEnergy}) can be found by perturbation
theory, in which $\delta U_{eff}=-\frac{\beta}{3} x_{Mn}({\bf
R})N_0(3/2)\bar{S}_z({\bf R})$ is a perturbation:

\begin{eqnarray}
\delta\Psi(R,\chi)= \sum_{(n_x,n_y)\neq
(0,0)}\frac{<\psi_{n,m}(R)|\delta
U_{eff}|\psi_{0,0}>}{\epsilon_{0,0}-\epsilon_{n,m}}\psi_{n,m}(R)|\uparrow>.
\nonumber
\\
\delta E_0(T)= \sum_{(n_x,n_y)\neq
(0,0)}\frac{|<\psi_{0,0}(R)|\delta
U_{eff}|\psi_{n,m}>|^2}{\epsilon_{0,0}-\epsilon_{n,m}}. \nonumber \\
\label{mpolaron2}
\end{eqnarray}
Now we can estimate the precision of the perturbation theory:
$\delta E_0/E_b$ is about a few $\%$ for the typical parameters of
the problem, which will be specified below. The precision is high
because of the orthogonality of the spatial harmonic-oscillator
functions $\psi_{n,n'}(r)$. For example, the nonzero matrix
element $<\psi_{2,0}(R)|\delta U_{eff}|\psi_{0,0}>$ at zero
temperature is about $1~meV$ and the perturbation-theory parameter
$<\psi_{n,m}(R)|\delta
U_{eff}|\psi_{0,0}>/(\hbar\omega_0)\sim0.02$. With increasing
temperature the parameter $<\psi_{n,m}(R)|\delta
U_{eff}|\psi_{0,0}>/(\hbar\omega_0)$ decreases and the precision
of the perturbation theory becomes further improved. The above
estimates tell us that the perturbation theory with respect to the
potential $\delta U_{eff}$ provides us with reliable results. We
also note that our results are in agreement with
ref.~\cite{Bhattacharjee0}.

The ferromagnetic state of a QD depends on the spatial
distribution of Mn ions, which is given by the crystal growth
technology. Diffusion of Mn ions during QD growth is driven by
strain and composition inhomogeneity and can lead to a strongly
non-uniform distribution of Mn ions in the system. For example, a
study of the $InMnAs$ system \cite{InMnAs}  has shown that the Mn
ions during the growth process mostly substitute for the In atoms
inside a QD. Here we are going to use a simple model
(fig.~\ref{fig1}) in which the Mn-doped region forms a disk of
radius $R_{Mn}$ inside a QD: $x_{Mn}(\rho)=x_{Mn}$ for
$\rho<R_{Mn}$ and $0$ otherwise. Here $\rho$ is the distance to
the center of a QD in the 2D plane. It follows from
eq.~\ref{MPolaronBEnergy} that the binding energy of a MP and the
"robustness" of a magnetic state depend on the overlap between the
Mn distribution $x_{Mn}(R)$ and the wave function of the hole.
Since both of them can be tailored and controlled by the growth
process the ferromagnetic state can be artificially designed.

\begin{figure}[tbp]
\includegraphics*[width=0.6\linewidth, angle=90]{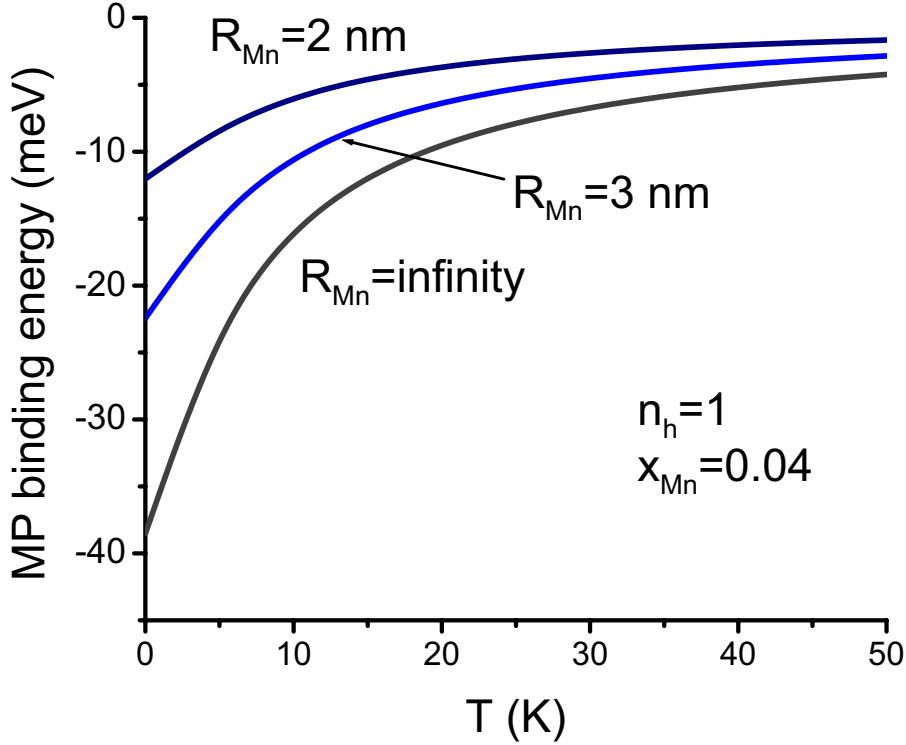}
\caption{Calculated energy of a MP with one hole for different Mn
distributions;  $x_{Mn}=0.04$ and  $R_{Mn}=2,3~nm$, and $\infty$.}
\label{fig2}
\end{figure}

\begin{figure}[tbp]
\includegraphics*[width=0.6\linewidth, angle=90]{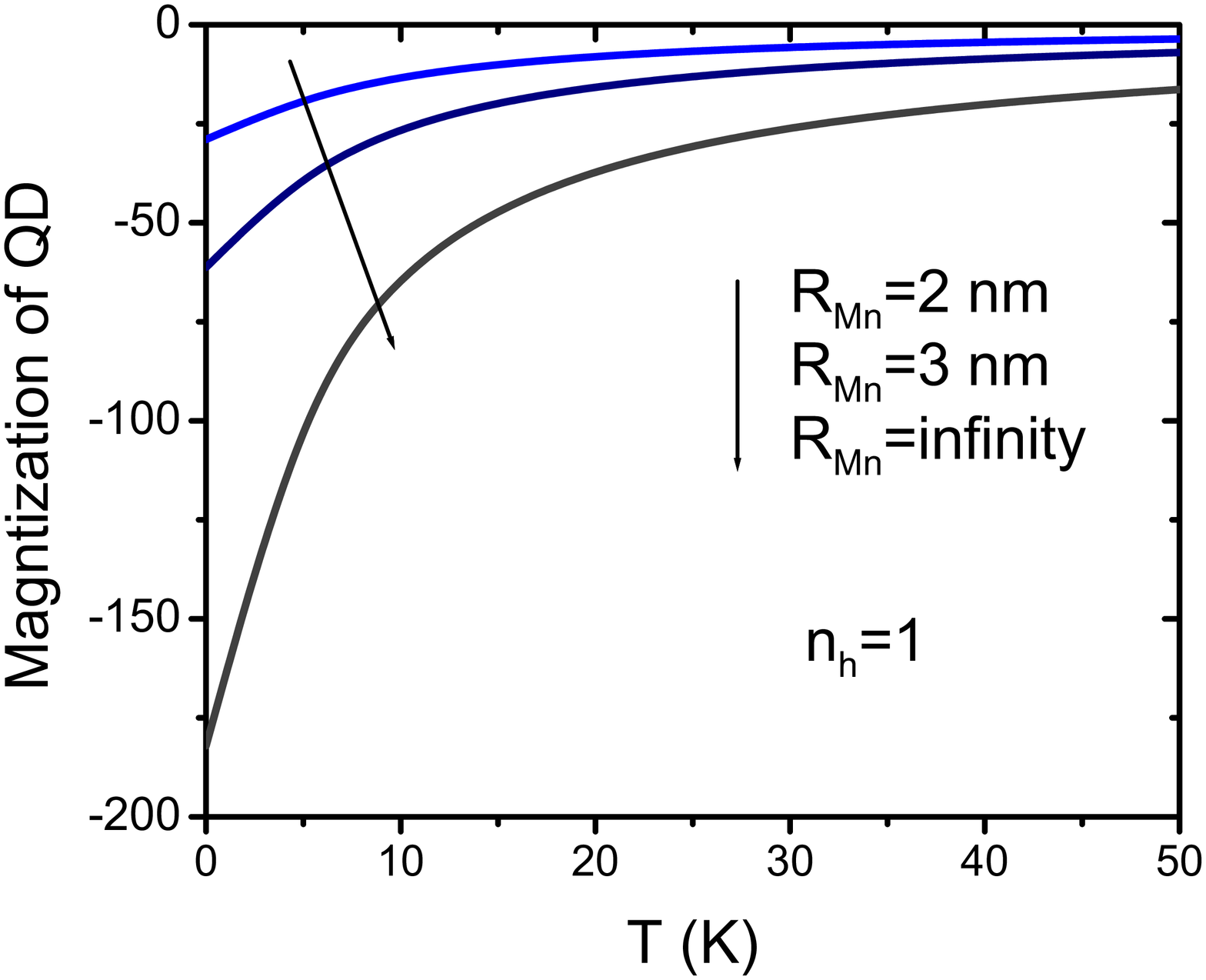}
\caption{Calculated magnetization of Mn ions as a function of
temperature for the QD with one hole and different Mn
distributions; $x_{Mn}=0.04$ and $R_{Mn}=2,3~nm$, and $\infty$.}
\label{fig3}
\end{figure}

Fig.~\ref{fig2} shows the calculated energy of a MP with one hole
for different Mn distributions, $R_{Mn}=2$, $3~nm$ and $\infty$.
The corresponding Mn magnetization is shown in fig.~\ref{fig3}.
The following parameters of a magnetic semiconductor have been
used: $x_{Mn}=0.04$, $\beta N_0=-1.3~eV$,
$T_0=3.6~K$, and $N_0=15~nm^{-3}$. The QD geometrical parameters
were chosen as follows: $l_0=4~nm$ and $L=2.5~nm$.
The above material parameters represent a $CdMnTe$ QD.

\section{Anisotropy of binding energy and magnetization}

\begin{figure}[tbp]
\includegraphics*[width=0.4\linewidth,angle=90]{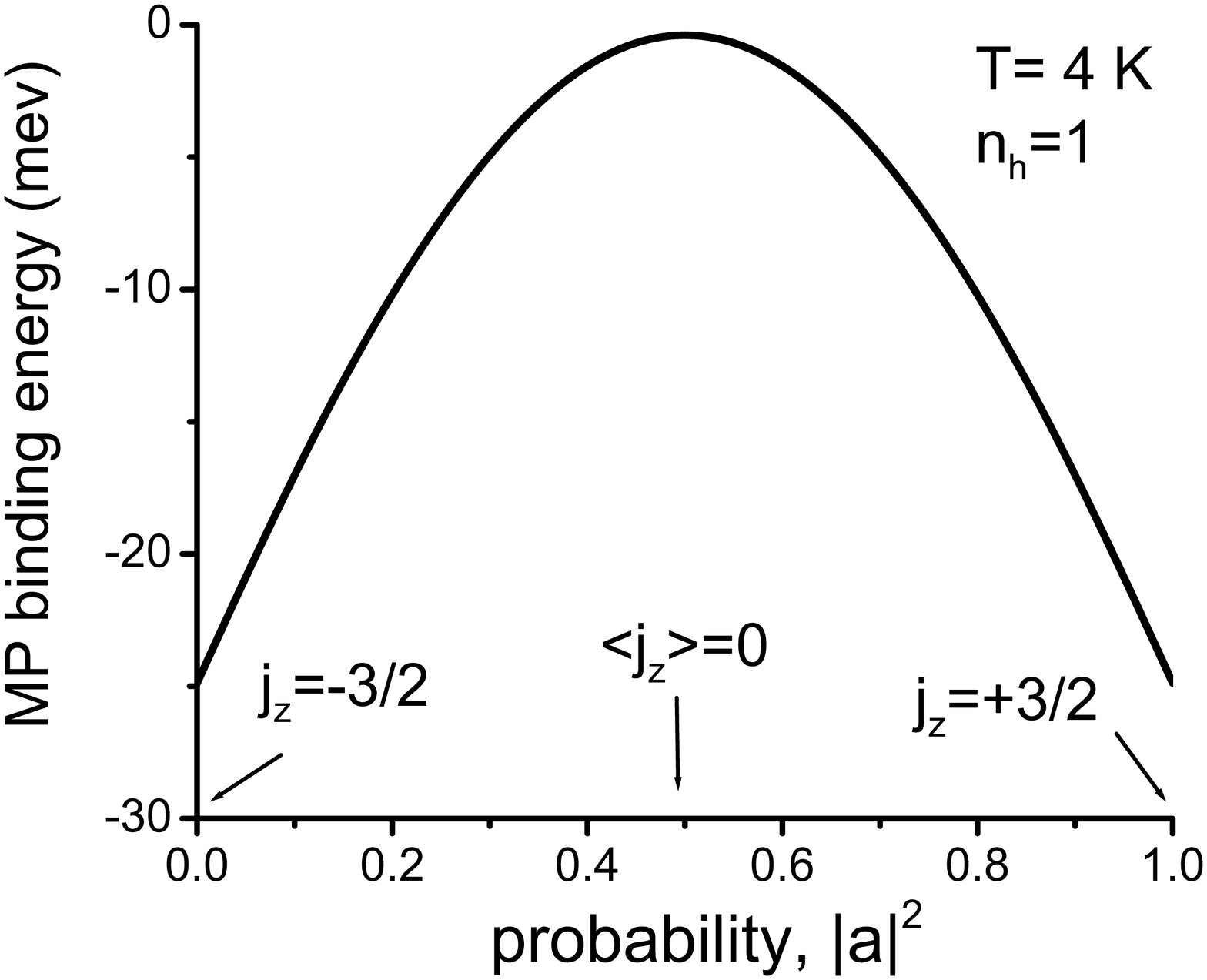}
\caption{Spin anisotropy of the MP energy as a function of the
probability to find the hole in the state $j_z=+3/2$;
$x_{Mn}=0.04$ and $R_{Mn}=\infty$.} \label{fig4}
\end{figure}

The MP state of a hole in a disk-shaped QD is strongly anisotropic
due to the valence band structure \cite{Mn-electron}. This
magnetic anisotropy comes from the heavy-light hole splitting in
the valence band and reveals itself in the last term in the
operator (\ref{Hamiltonian}). The general solution of the one-hole
problem can be written as

\begin{eqnarray}
\Psi=\psi_{0,0}(a|\uparrow>+b|\downarrow>), \label{mpolaronAnis1}
\end{eqnarray}
where $|a|^2+|b|^2=1$. Then the MP binding energy takes the form:

\begin{eqnarray}
E_{b}(T)=-\frac{\beta}{3}\int_R{d^3R [\psi_{0,0}^2({\bf R})
x_{Mn}({\bf
R})N_0\frac{3}{2}(2|a|^2-1)SB_S(\frac{\beta/3\psi_{0,0}(R)^2\frac{3}{2}[2|a|^2-1]}{k_B(T+T_0)})]},
\label{MPolaronBEnergyAniz}
\end{eqnarray}
where $0\leq|a|\leq1$. Figure~\ref{fig4} shows the MP binding
energy as a function of $|a|$. For the cases $|a|=1$ and $|b|=1$,
the binding energy magnitude is maximum; for $|a|=1/\sqrt{2}$ it
equals to zero. Therefore, the ground state of a MP corresponds to
the pure states $|\uparrow>$ or $|\downarrow>$. This anisotropy
likely plays an important role in optical experiments with
excitons trapped in semi-magnetic QDs \cite{Cincinnati}. In such
experiments, an optically-created electron-hole pair rapidly
relaxes to its ground state resulting in the formation of a MP
with the spin parallel (or antiparallel) to the growth axis.

\section{Few particle states}

\begin{figure}[tbp]
\includegraphics*[width=0.6\linewidth,angle=90]{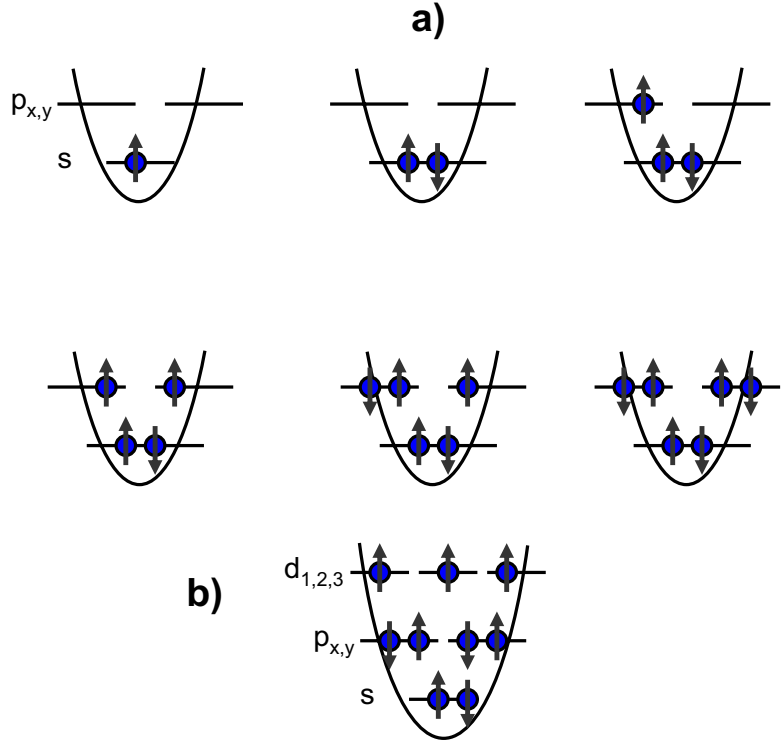}
\caption{(a) Ground-state configuration of a cylindrically
symmetric QD with few holes and without Mn ions; the state with
$n_h=4$ is constructed according to Hund's rule. These
configurations remain ground states if the Mn-hole interaction is
weak or the QD confinement is strong enough. (b) Ground-state
configuration for $n_h=9$ in the regime of Hund's rule; again this
configuration represents a ground state if the Mn-hole interaction
is not very strong. } \label{fig5}
\end{figure}

In the next step, we study the QD magnetization in the presence of
few carriers which can be loaded from the metal back contact in a
voltage-tunable transistor structure (fig.~\ref{fig1}b). We will
sequentially consider the few first charged states of a QD
starting from $n_h=1$ (fig.~\ref{fig5}). The Coulomb interaction
will be treated within perturbation theory which is valid for QDs
with a strong confinement; namely, we will assume that the
quantization energy of QD, $\hbar\omega_0$, is larger than the
characteristic parameter of Coulomb interaction between particles,
$E_{Coul}=e^2/(\epsilon~l)$. For the QD parameters specified
above, we obtain: $\hbar\omega_0\approx47~meV$ and
$E_{Coul}\approx29~meV$. This simplified perturbation approach is
very convenient and was successfully used for description of
experimental data in several publications
\cite{QDVoltage1,QDVoltage3,Richard-Nature,Nature2004}. Note that
the parameters of the QDs studied in \cite{QDVoltage1,QDVoltage3}
are close to those used here.

In the perturbation approach, we will neglect the Coulomb-induced
mixing between shells; at the same time, we will calculate exactly
the Coulomb-induced mixing within shells by diagonalizing the
corresponding matrix. Thus, the Coulomb correlations will play a
very important role for certain states, such as the Hund's states
with two particles in the $p$-shell ($n_h=4$) and three holes in
the $d$-shell (n=9).

For some derivations, it will be convenient to treat this problem
using the second quantization approach. In this approach, the
z-component of the total angular momentum of the hole subsystem is
given by

\begin{eqnarray}
\hat{j}_{z,tot}(R)=
\sum_{\gamma,j_z=\mp3/2}j_z|\psi_{\gamma}(R)|^2\hat{c}^+_{\gamma,j_z}\hat{c}_{\gamma,j_z},
\label{MomentumTotalSCa}
\end{eqnarray}
where $\hat{c}^+_{\gamma,j_z}$ is the creation operator for the
single-particle state $(\gamma,j_z)$. Here $\gamma$ stands for the
pair of orbital quantum numbers; $\gamma$ may be $n_x,n_y$ or
$n,m_z$ depending on the choice of wave functions; here $n$ and
$m_z$ are the radial quantum number and the orbital angular
momentum, respectively. For the few-particle wave functions, we
will employ the following notations:
$|s_{\uparrow},s_{\downarrow};p_{1,\uparrow},p_{1,\downarrow};p_{2,\uparrow},p_{2,\downarrow};
d_{1,\uparrow},d_{1,\downarrow};d_{2,\uparrow},d_{2,\downarrow};d_{3,\uparrow},d_{3,\downarrow};...>$,
where $s_{\uparrow(\downarrow)}$, $p_{i,\uparrow(\downarrow)}$,
and $d_{k,\uparrow(\downarrow)}$ are the occupation numbers for
the s-, p-, and d- states with the corresponding spins. These
occupation numbers can be either $0$ or $1$. For the indices $i$
and $k$, $i=1,2$ and $k=1,2,3$.

\section{Quantum dot with two holes}

\begin{figure}[tbp]
\includegraphics*[width=0.6\linewidth,angle=0]{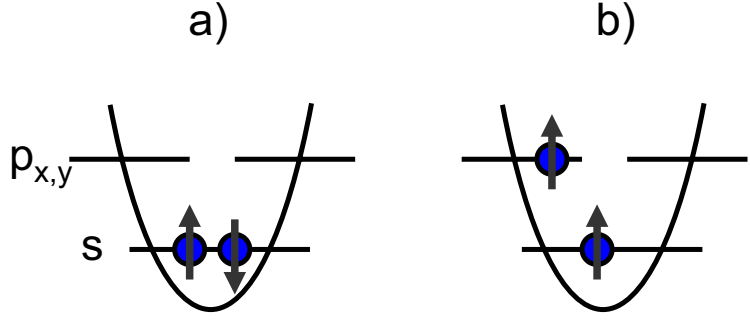}
\caption{Ground-state configurations of a QD with cylindrical
symmetry and two holes. The configuration (a) is realized in the
case of a relatively weak Mn-hole interaction (high temperature or
small Mn density) and the state (b) corresponds to the system with
a strong Mn-hole interaction (low temperature or high Mn
density).} \label{fig6}
\end{figure}

If the QD is occupied by two particles ($n_h=2$), the spin and
spatial variables in the two-hole wave function can be separated.
Then, for a QD with strong confinement, it becomes obvious that
the z-component of the total hole spin in the ground state is zero
($j_{z,tot}=0$) (fig.~\ref{fig6}a) and therefore the exchange
interaction with the Mn subsystem vanishes. However, if the
confinement is not strong enough, the ground state can change with
temperature. Figure \ref{fig6}b shows the hole configuration which
can become a ground state if the Mn-hole interaction is strong
enough. This state has the total spin $j_{z,tot}=2*(3/2)$ and a
non-zero binding energy in the presence of the Mn subsystem. The
ground state transition occurs when

\begin{eqnarray}
E_1-E_0=|E_{b}^{n_h=2}|. \label{TransEnergy}
\end{eqnarray}
Here $E_0$ and $E_1$ are the hole energies of the states (a) and
(b) in fig.~\ref{fig6}. The hole energies $E_{0(1)}$ also include
the contributions from the Coulomb interaction. $E_{b}^{n_h=2}$ is
the binding energy for the configuration (b) in fig.~\ref{fig6}.

The Hamiltonian of the system is now written as:

\begin{eqnarray}
\hat{H}_{n_h=2}=\sum_{i=1,2}[\hat{T}_i+U({\bf
R}_i)]-\frac{\beta}{3} x_{Mn}({\bf R})N_0 \hat{j}_{z,tot}
\bar{S}_z({\bf R}_i)+U_{Coul}, \label{Hamiltonian2holes}
\end{eqnarray}
where $\hat{j}_{z,tot}$ is the z-component of the total momentum
(spin) of holes, $\hat{T}_i$ is the kinetic energy of the
$i$-hole, and $U_{Coul}$ is the Coulomb interaction. For the
Coulomb potential, we use the usual formula:
$e^2/\epsilon|R_1-R_2|$, where $\epsilon$ is the dielectric
constant.

\begin{figure}[tbp]
\includegraphics*[width=0.6\linewidth, angle=90]{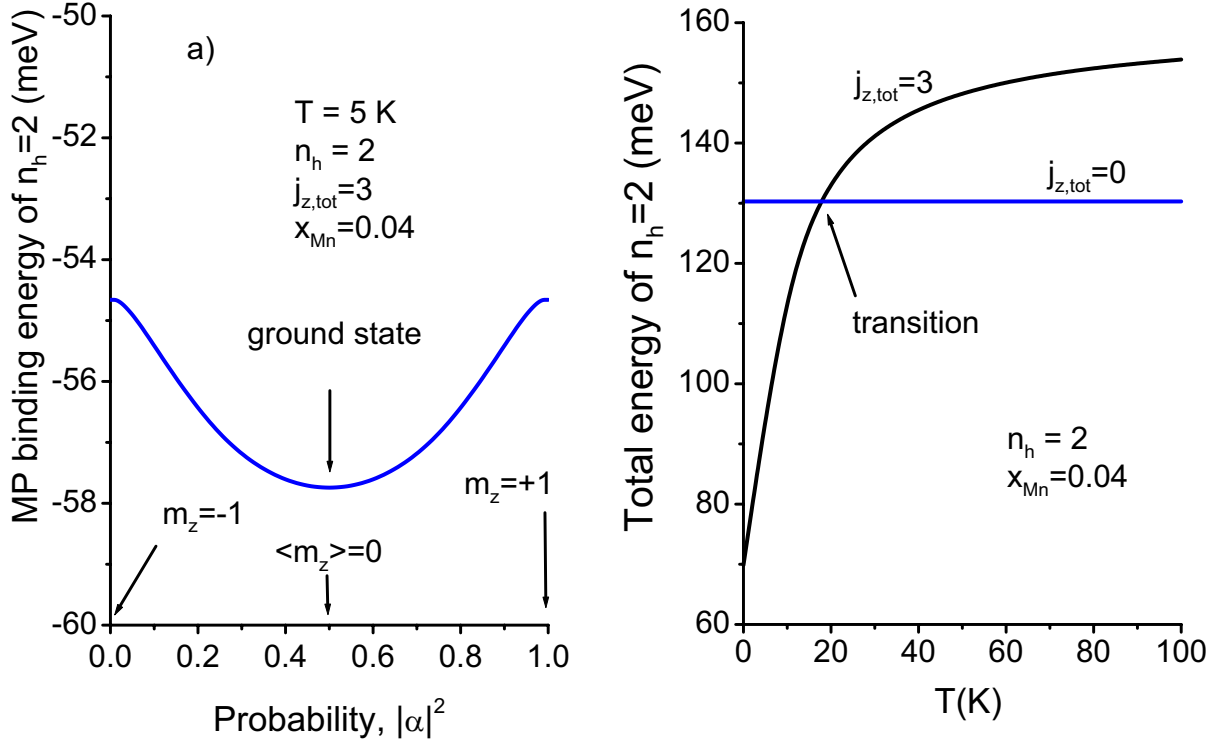}
\caption{(a) Orbital anisotropy of the MP energy with $n_h=2$ and
$j_{z,tot}=3$. The minimum energy corresponds to the state with
$|\alpha|^2=1/2$ in which the angular part of the spatial
probability distribution is most inhomogeneous, i.e.
$\psi_p^2(\phi)\propto cos(\phi+\phi_0)^2$. (b) The energy of the
states $j_{z,tot}=3$ and $j_{z,tot}=0$ (figs.~6a and b) as a
function of temperature; the ground state of the system changes
with temperature. $R_{Mn}=\infty$ and $x_{eff}=0.04$} \label{fig7}
\end{figure}

Regarding the Coulomb interaction in eq.~\ref{Hamiltonian2holes},
we will treat it again as a perturbation assuming a
strongly-confined QD
\cite{QDVoltage1,QDVoltage3,Richard-Nature,Nature2004}. Then, we
obtain for the energies of the states:
$E_0=2\hbar\omega_0+U_{ss}^{dir}$ and
$E_1=3\hbar\omega_0+U_{sp}^{dir}-U_{sp}^{exc}$, where
$U_{\alpha,\alpha'}^{dir}$ and $U_{\alpha,\alpha'}^{exc}$ are the
direct and exchange Coulomb elements, respectively. The Coulomb
matrix elements $U_{sp}^{dir(exc)}$ do not depend on a particular
choice of the single-particle wave functions for the p-shell. For
the state (b) in fig.~\ref{fig6}, the operator
(\ref{MomentumTotalSCa}) includes two terms:

\begin{eqnarray}
\hat{j}_{z,tot}(R)=\frac{3}{2}(|\psi_{0,0}(R)|^2\hat{c}^+_{s,\uparrow}\hat{c}_{s,\uparrow}+
|\psi_{p}(R)|^2\hat{c}^+_{p,\uparrow}\hat{c}_{p,\uparrow}),
\label{MomentumTotalSC}
\end{eqnarray}
where the p-state orbital wave function $\psi_{p}$ can be written
as:

\begin{eqnarray}
\psi_p=\alpha\psi_{+}+\beta\psi_{-}; \label{mpolaronAnis2}
\end{eqnarray}
here $\psi_{+(-)}(\phi)\propto e^{\pm\phi}$ and
$|\alpha|^2+|\beta|^2=1$, where ${\bf r}=(\rho,\phi)$. The indices
$+(-)$ correspond to the wave functions with the orbital angular
momentum $m_z=+1(-1)$, respectively. The binding energy of the MP
with two holes

\begin{eqnarray}
E_{b}(T,\alpha)=-\frac{\beta}{3}\int_R{d^3R [F_2(R) x_{Mn}({\bf
R})N_0\frac{3}{2}SB_S(\frac{\beta/3F_2(R)\frac{3}{2}}{k_B(T+T_0)})]},
\label{MPolaronBEnergy2holes}
\end{eqnarray}
where $F_2(R)=|\psi_{0,0}(R)|^2+|\psi_{p}(R)|^2$ is the particle
density in the system. The binding energy now depends on the
parameter $\alpha$. We find numerically that the ground state is
realized for $|\alpha|^2=1/2$. This ground state corresponds to
the wave function with the most inhomogeneous p-orbital as a
function of $\phi$: $\psi_p\propto cos(\phi+\phi_0)$, where
$\phi_0$ is an arbitrary phase. It can be understood as follows:
in the ground state, the spatial wave function of holes should be
most localized because a strongly localized hole can better
control the Mn spins and more strongly lower the total energy.
This case resembles somewhat a self-trapped MP in the systems with
translation invariance \cite{MPselftrapped}. We note that the
ground state of MP is degenerate since the phase $\phi_0$ is
arbitrary in a system with cylindric symmetry. However, if the QD
confinement is anisotropic, the phase $\phi_0$ will be fixed by
the anisotropy of the QD potential.

In fig.~\ref{fig7}a, we show the MP binding energy as a function
of $|\alpha|^2$. When $|\alpha|^2=1$ or $0$, the p-hole state has
the orbital angular momentum $m_z=+1$ or $m_z=-1$, when
$|\alpha|^2=1/2$, it has linear polarization. For a QD with the
parameters $m_{hh}=0.1 m_0$, $\hbar\omega_0=47.3~meV$, and
$\epsilon=12.5$, the Mn-hole system demonstrates a critical
temperature at which the ground state changes from the state
$j_{z,tot}=3$ to the state with $j_{z,tot}=0$ (fig.~\ref{fig7}b).
This critical temperature corresponds to the solution of
eq.~\ref{TransEnergy}. The existence of this transition depends on
the particular choice of parameters of a QD. For example, if the
density of Mn impurities is small enough, the ground state will be
always $j_{z,tot}=0$ and the transition will not occur at all.

\section{Quantum dot with a smaller $Mn$ density and $n_h=1-6$}

\begin{figure}[tbp]
\includegraphics*[width=0.6\linewidth,angle=90]{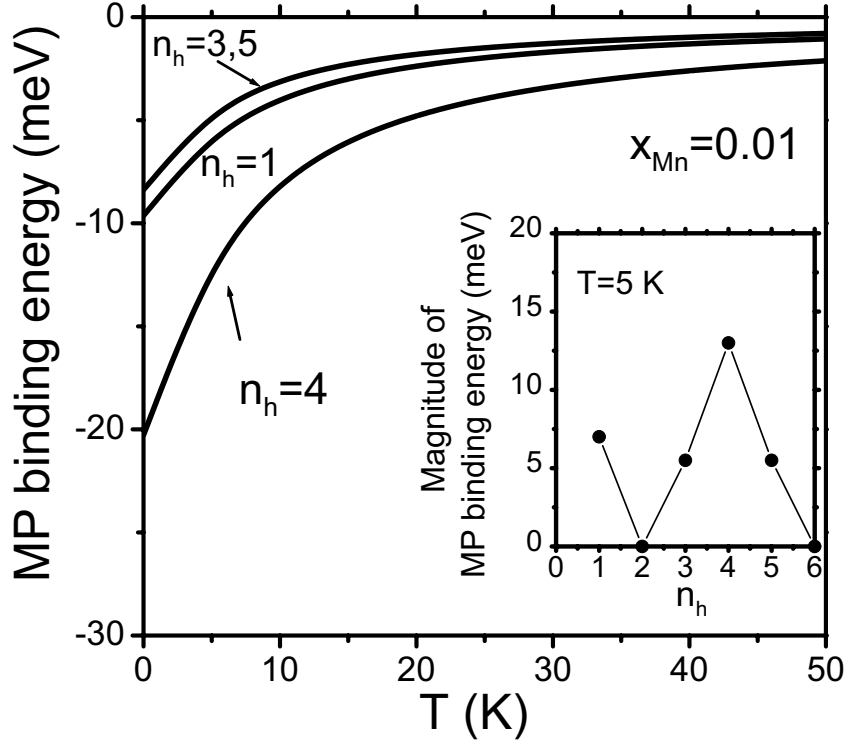}
\caption{Calculated binding energy of a MP as a function of
temperature for various charged states of the QD: $n_h=1,3,4$ and
$5$. The binding for the states with completely filled shells
($n_h=2,6$) is zero in our approach; $R_{Mn}=\infty$.}
\label{fig8}
\end{figure}

We now consider a QD with $n_h$ ranging from $1$ to $6$. As was
pointed out above Coulomb correlations will become very important
for certain states. In addition, we will assume that the Mn
density is a few times smaller than in the previous calculations
and the ground-state transitions of the type shown in
fig.~\ref{fig7} do not occur. Therefore, the ground state
configurations coincide with those in a non-magnetic QD
(fig.~\ref{fig5}). The calculated binding energies for the lower
doping $x_{eff}=0.01$ are shown in fig.~\ref{fig8}.

We now proceed to the case of three holes ($n_h=3$). In this case,
the ground state has two holes in the s-shell and one in the
p-shell. The closest excited state has one hole in the s-shell and
two spin-polarized holes in the p-shell ($j_{z,tot}=3*(3/2)$).
Despite the Mn-hole interaction, the configuration
$j_{z,tot}=3*(3/2)$ remains an excited state since the additional
MP binding energy in this state is about $20~meV$ and
significantly lower than the quantization energy $\hbar\omega_0$
for our parameters. In the ground state with $n_h=3$, a nonzero
angular momentum of the hole system comes from the unpaired hole
in the p-state. The result for the MP binding energy is given by
eq.~\ref{MPolaronBEnergy2holes} with $F_2(R)=|\psi_{p}(R)|^2$.
Again the ground state is anisotropic due to the degeneracy of the
p-shell.

\begin{figure}[tbp]
\includegraphics*[width=0.6\linewidth,angle=90]{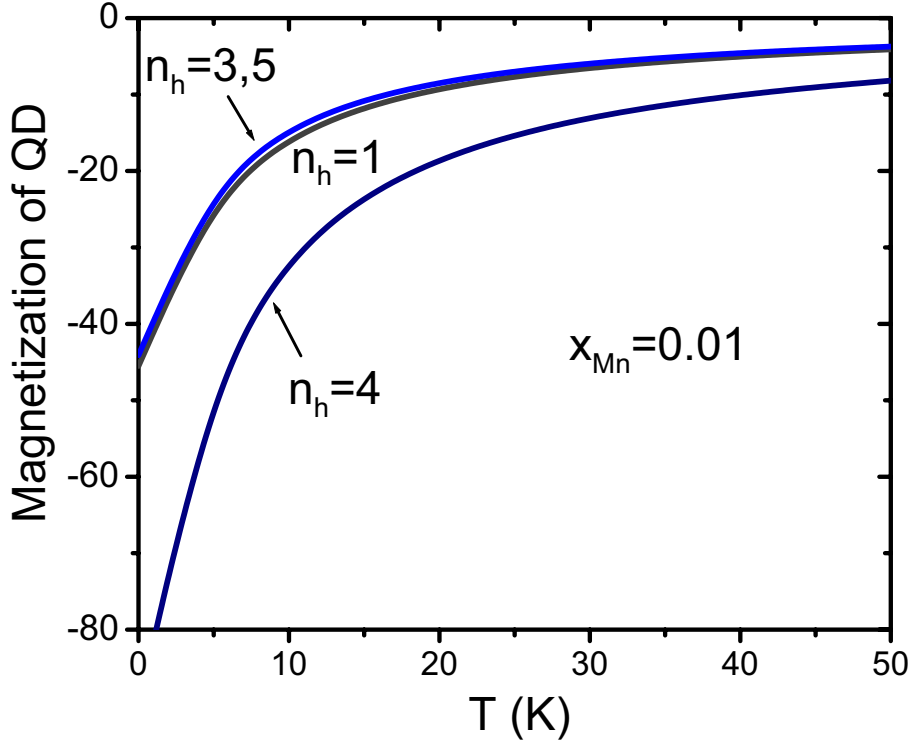}
\caption{Calculated magnetization of Mn ions as a function of
temperature for various charged states of QD: $n_h=1,3,4$ and $5$.
The magnetization for the states with completely filled shells
($n_h=2,6$) is zero in our approach; $x_{eff}=0.01$ and
$R_{Mn}=\infty~nm$.} \label{fig9}
\end{figure}

To describe the ground state of a QD with $n_h=4$, we should apply
Hund's rule. In the state $n_h=4$, the s-shell is completely
filled, whereas the p-shell is occupied by two holes. According to
Hund's rule, the ground state of the many-particle system without
the Mn subsystem should have two particles with parallel spins in
the upper shell and the maximum spin $3$. The corresponding wave
function should be found by mixing the Slater determinants related
to the p-shell and diagonalizing the Coulomb matrix. The states
with the smallest energy form a triplet:

\begin{eqnarray}
|1,1;1,0;1,0>, \hskip 0.4cm |1,1;0,1;0,1>, \hskip 0.4cm
\frac{|1,1;0,1;1,0>+|1,1;1,0;0,1>}{\sqrt{2}}. \label{fourholes}
\end{eqnarray}
For the case $n_h=4$, the first excited states within the p-shell
are $|1,1;1,1;0,0>$ and $|1,1;0,0;1,1>$. The above wave functions
should be represented by Slater determinants composed of the
orbitals $\psi_s$ and $\psi_{\pm}$. The state with the largest
energy has the configuration:
$\frac{|1,1;0,1;1,0>-|1,1;1,0;0,1>}{\sqrt{2}}$. The energies of
the above states are given by: $E_0-U_{pp}^{exc}$, $E_0$, and
$E_0+U_{pp}^{exc}$, where $U_{pp}^{exc}$ is the exchange integral
and $E_0$ involves single-particle energies and some Coulomb
interactions. In the absence of the Mn subsystem, the difference
of energy between the ground state ($j_{z,tot}=3$) and the first
excited states ($j_{z,tot}=0$) is equal to the exchange energy
between p-states: $U^{exc}_{pp}=(e^2/\epsilon)\int dR_1^3dR_2^3
\frac{\psi_+(R_1)^*\psi_-(R_2)^*\psi_+(R_2)\psi_-(R_1)}{|R_1-R_2|}$.
Moreover, the energy of the state $\Psi_0=|1,1;1,0;1,0>$ will be
lowered due to interaction with Mn spins and this lowering can be
very significant. The reason is the 2-fold increase of the total
hole spin:

\begin{eqnarray}
<\Psi_0|\hat{j}_{z,tot}(R)|\Psi_0>=2\frac{3}{2}|\psi_+(R)|^2.
\label{MomentumTotalSCME}
\end{eqnarray}
The MP binding energy for this state is

\begin{eqnarray}
E_{b,n_h=4}(T)=-\frac{\beta}{3}\int_R{d^3R [|\psi_{+}(R)|^2
x_{Mn}({\bf R})3
N_0SB_S(\frac{\beta|\psi_{+}(R)|^2}{k_B(T+T_0)})]}.
\label{MPolaronBEnergyPshell2}
\end{eqnarray}
We note that the increase of the binding in
eq.~\ref{MPolaronBEnergyPshell2} comes from the factors 2 before
and inside the Brillouin function. Also, the MP energy
(eq.~\ref{MPolaronBEnergyPshell2}) in the case of $n_h=4$ does not
depend on the particular choice of the single particle functions;
in the coordinate representation, the p-electrons are described
with the anti-symmetric wave function: $\Psi_0\propto
sin(\phi_1-\phi_2)|\uparrow_1>|\uparrow_2>$. The magnitude of the
calculated MP binding energy demonstrates a strong increase (about
two times) (fig.~\ref{fig8}). The Mn magnetization also increases
for the case $n_h=4$ (fig.~\ref{fig9}).

The ground state $n_h=5$ is again non-uniform as a function of
angle $\phi$:

\begin{eqnarray}
\Psi_0^{n_h=5}=\alpha|1,1;1,1;1,0>+\beta|1,1;1,0;1,1>.
\label{fiveholes}
\end{eqnarray}
When we ignore inter-shell mixing, the results for $n_h=5$ become
similar to those for $n_h=3$. The magnetization comes from one
unpaired p-hole:

\begin{eqnarray}
<\Psi_0|\hat{j}_{z,tot}(R)|\Psi_0>=+\frac{3}{2}|\alpha\psi_{+}+\beta\psi_{-}|^2.
\label{MomentumTotalSpin5holes}
\end{eqnarray}
The lowest MP energy is obtained for the "linearly-polarized"
state of holes ($\alpha=1/\sqrt{2}$) as in the case of $n_h=3$
(fig.~\ref{fig3}).

The state $n_h=6$ has completely filled s- and p-shells and
spontaneous Mn magnetization does not appear for QDs with a
relatively weak Mn-hole interaction (the inter-shell mixing is
neglected).

\section{formal self-consistent solution for a few particle quantum dot:
toward the ferromagnetic phase transition.}

The system of mobile carriers can undergo a ferromagnetic
transition if the Mn carrier interaction is strong enough and
exceeds the anti-ferromagnetic interaction between ions. The Curie
temperature of Zener ferromagnetism is given by the density of
states of mobile carriers, the density of Mn ions, and the Mn-hole
interaction constant \cite{Zener}:

\begin{eqnarray}
T_{Curie}\propto S(S+1)x_{eff}N_0\rho(E_F)\beta^2/k_B,
 \label{TCurie}
\end{eqnarray}
where $\rho(E_F)$ is the density of states at the Fermi level. It
is know that the MP state in a QD does not undergo the phase
transition: the spontaneous polarization in the MP state simply
decreases with temperature \cite{MPBulk}. This is due to the fact
that the system contains just one or few carriers.

We now consider a formal self-consistent solution of the mean
field theory in a manner similar to the Zener theory. For $n_h=1$,
we employ eq.~\ref{Brillion} and obtain a self-consistent integral
equation

\begin{eqnarray}
\bar{S}_z({\bf R})=S B_S(\frac{\beta/3<j_z>_T}{k_B(T+T_0)}),
\label{Brillion3}
\end{eqnarray}
where the $<j_z>_T$ is now averaged over the states of a hole:

\begin{eqnarray}
<j_z>_T=\frac{3}{2}\psi_{0,0}^2\frac{e^{-\frac{\Delta_1}{k_BT}}-
e^{\frac{\Delta_1}{k_BT}}}{Z_s(T)}, \label{Brillion4}
\end{eqnarray}
where the quantity
\begin{eqnarray}
\Delta_1=-\frac{\beta}{3}\frac{3}{2}\int
d^3R[\psi_{0,0}^2x_{eff}N_0\bar{S}_z({\bf R})]  \label{Delta1}
\end{eqnarray}
plays the role of binding energy for the hole state $+3/2$;
$Z_s=e^{-\frac{\Delta_1}{k_BT}}+e^{\frac{\Delta_1}{k_BT}}$ is the
partition sum for the s-shell. The inter-shell mixing is again
ignored. Assuming a homogeneous spatial Mn distribution and
integrating eq.~\ref{Brillion3}, we obtain:

\begin{eqnarray}
\Delta_1=-\frac{\beta}{3}\frac{3}{2}S x_{eff} N_0 f_1(\Delta_1),
\label{Delta2}
\end{eqnarray}
where $f_1(\Delta_1)=\int
B_S(\frac{\beta/3<j_z>_T}{k_B(T+T_0)})d^3R$. This simple equation
leads to the critical behavior: a solution with nonzero
magnetization ($\Delta_1>0$) exists if $T<T_{crit}$ where

\begin{eqnarray}
k_BT_{crit}^{n_h=1}=\sqrt{
(\frac{\beta}{3})^2j_h^2\frac{S(S+1)}{3}x_{eff}N_0\int
d^3R\psi_{0,0}^4}-k_BT_0\label{Tcrit}
\end{eqnarray}

This critical behavior occurs in the regime of strong fluctuations
in the MP state and obviously is incorrect. It is known that MPs
do not exhibit a critical behavior. However, it seems to be
interesting to compute how this formal self-consistent solution
develops with increasing number of holes in the QD. First, the
self-consistent approach becomes more reliable with increasing the
number of particles. Second, eq.~\ref{Tcrit} gives a useful
estimate of the characteristic temperature of spontaneous
magnetization for the MP states in a QD.

\begin{figure}[tbp]
\includegraphics*[width=0.6\linewidth,angle=0]{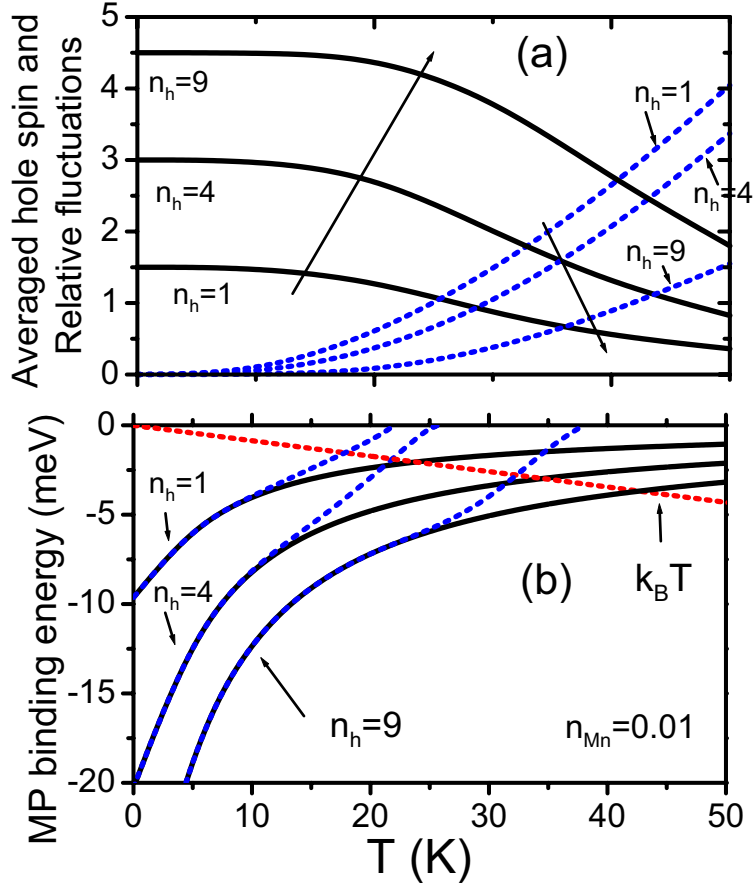}
\caption{(a) Calculated average spin of the hole subsystem (solid
curves) and its relative fluctuations (dashed lines) for the
states $n_h=1,4$, and $9$. One can see that the ferromagnetic
state becomes more stable with increasing number of holes. (b) The
solid curves show the calculated binding energy of a MP as a
function of temperature for three charged states of a QD
($n_h=1,3$, and $9$) with maximum binding energy; $x_{Mn}=0.01$
and $R_{Mn}=\infty$. The dashed curves show the results of
self-consistent mean-field theory which demonstrates critical
temperature. The dashed straight line is the thermal energy. }
\label{fig10}
\end{figure}

For the cases of $n_h=4$ and $n_h=9$, we have found similar
self-consistent solutions for the energy and magnetization by
solving the corresponding nonlinear equations. For example, in the
case of $n_h=9$, the ground-state configuration is constructed
according to Hund's rule (fig.~\ref{fig5}b) and the
self-consistent problem is reduced to a system of two nonlinear
equations of two variables: $x_1=\int
d^3R[\psi_{d,0}^2\bar{S}_z({\bf R})]$ and $x_2=\int
d^3R[\psi_{d,+}^2\bar{S}_z({\bf R})]$, where $\psi_{d,0}$ and
$\psi_{d,+}$ are the d-orbitals with $m_z=0$ and $+2$,
respectively. Again we ignored inter-shell mixing and diagonalized
the Coulomb matrix inside the p- and d-shells. Then, using the
obtained energies, we constructed the partition sums for the p-
and d-shell. In figure \ref{fig10}b, we show by dashed lines the
averaged energies of MPs calculated within the self-consistent
approach. In the same figure, the solutions without the
self-consistent averaging of the spin inside the Brillouin
function are shown as solid curves.

Equation (\ref{Tcrit}) gives a characteristic temperature at which
spontaneous Mn polarization exists in a QD. From fig.~\ref{fig10},
we see that the critical temperature increases with number of
holes. However, this increase is not very strong. The reason is
that the p- and d-shells have several states and some of these
states have opposite spins or no spin. These states contribute to
the partition sum and therefore effectively reduce the
magnetization.

It is interesting to compare the typical temperature of MPs with
that of ferromagnetic phase transitions in bulk. For CdMnTe, it
was suggested that the Curie temperature for the highest hole
densities can be as high as a few Kelvins \cite{estimateTC}. In a
QD with about 20 Mn ions considered here ($x_{eff}=0.01$,
$l=4~nm$, and $L_z=2.5~nm$), the typical temperature given by
eq.~\ref{Tcrit} is about $22~K$, an order of magnitude larger than
that in bulk. This is also consistent with previous papers on MPs
\cite{Bhattacharjee,Brey}. We can also note the important
differences between the above equations for $T_{Curie}$ and
$T_{crit}$. The critical temperature is proportional to the first
power of the interaction $\beta$ and depends on the QD
localization length ($T_{crit}\sim l^{-3/2}$).

\section{Fluctuations as a function of $n_h$}

To better understand the behavior of MPs at high temperature, we
now compute the relative fluctuations of the hole subsystem:

\begin{eqnarray}
\delta j_{z,tot}(T,n_h)=
\frac{\sqrt{<(\hat{j}_{z,tot}-<\hat{j}_{z,tot}>_T)^2>_T}}{<\hat{j}_{z,tot}>_T},
\label{Fluctuations}
\end{eqnarray}
where $<...>_T$ means thermal averaging over the many-particle
states assuming a given Mn spin distribution. For the Mn spin
distribution we will use eq.~\ref{Brillion} calculated for a
certain quantum state of the hole subsystem.

In the low-temperature regime, the relative fluctuations are much
weaker than the average spin of holes (fig.~\ref{fig10}). With
increasing temperature, the fluctuations grow and exceed
$<\hat{j}_{z,tot}>_T$. Figure \ref{fig10} shows clearly the
tendency of stabilization of the MP state with increasing number
of holes for the states $n_h=1,3,9$. This stabilization is
consistent with the increase of binding. Clearly, methods beyond
the mean-field theory are required to understand the mechanism of
crossover from the MP behavior to the Zener ferromagnetic phase
transition \cite{MPBulk2,fluct}.

\section{Capacitance of QD systems with magnetic polarons}

One efficient method to study quantum states of QDs is capacitance
spectroscopy. Such spectroscopy is typically performed at a
nonzero frequency $\omega$. The capacitance of the QD structure
(fig.~\ref{fig1}b) includes the contribution of charges trapped
inside the QD layer \cite{QDVoltage3}:

\begin{figure}[tbp]
\includegraphics*[width=0.6\linewidth,angle=90]{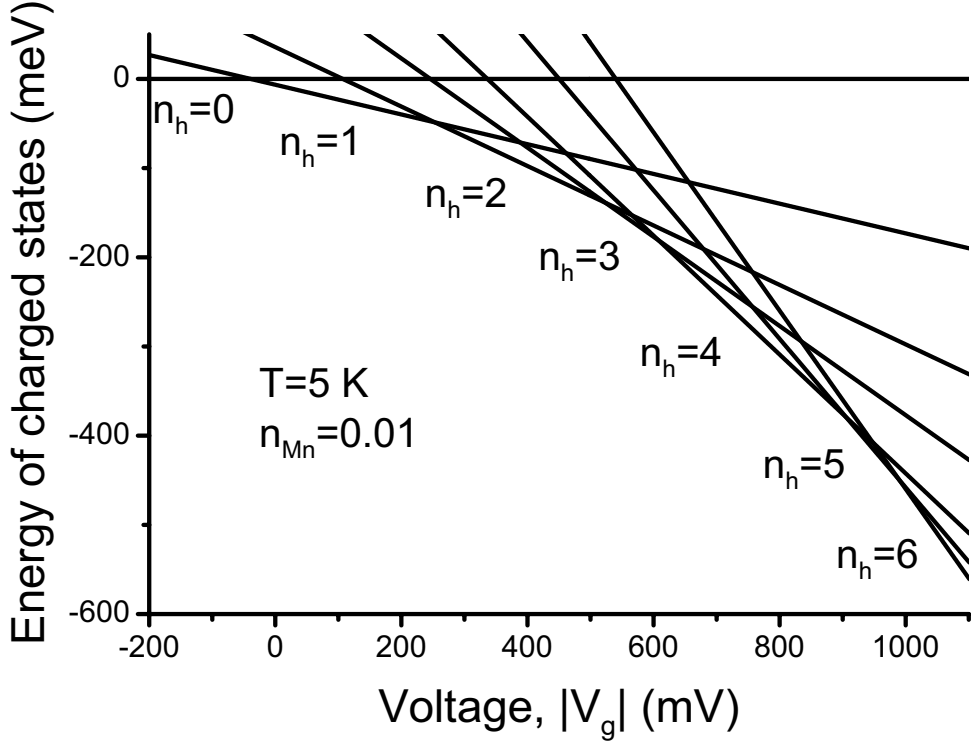}
\caption{(a) Calculated energies of different charged states in
the layer of semi-magnetic quantum dots with $x_{Mn}=0.01$;
$R_{Mn}=\infty$. The labels show the charged states of the
system.} \label{fig11}
\end{figure}

\begin{figure}[tbp]
\includegraphics*[width=0.6\linewidth,angle=90]{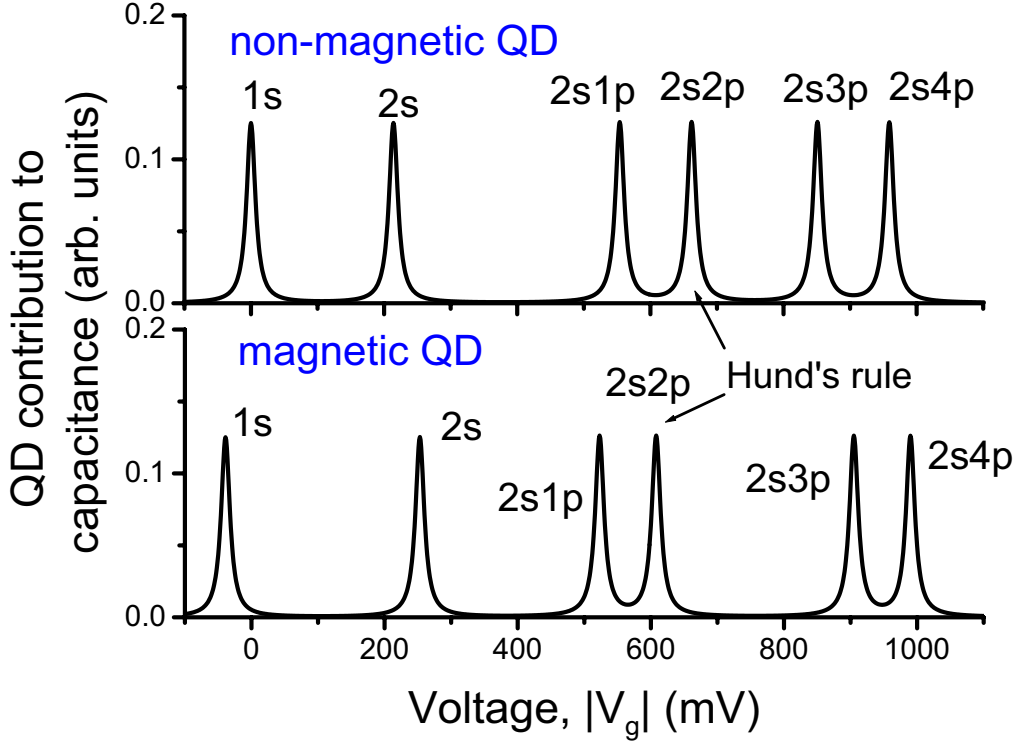}
\caption{(a) Calculated capacitance spectra for non-magnetic (a)
and magnetic (b) QDs; $x_{eff}=0.01$. The broadening of the peaks
was taken as $16~meV$.} \label{fig12}
\end{figure}

\begin{eqnarray}
\delta C=e^2 D(V_g), \label{CapacA}
\end{eqnarray}
where $V_g$ is the gate voltage and $D(V_g)$ is the effective
density of states in the QD layer. The latter is defined as

\begin{eqnarray}
D(V_g)=N_{dot}\frac{dn_h}{|e|d V_g}, \label{CapacB}
\end{eqnarray}
where $n_h$ is the number of holes trapped in a single QD and
$N_{dot}$ is the 2D density of QDs. This approach to the
capacitance is valid at low frequencies $\omega\tau\ll1$, where
$\tau$ represents both the tunneling time and the relaxation time
to form the ground MP state. In other words, this approach assumes
that the weakly coupled system "QD + metal contact" has a short
relaxation time and always remains in its ground state while the
gate voltage changes in time as $V_g+\delta V_g cos(\omega t)$
($V_g\gg\delta V_g$).

According to the simple model of a field-effect structure with a
QD layer \cite{QDVoltage3}, the energy of a single particle in the
QD is written as: $E_0(V_g)=E_{sp}+|e|\gamma V_g$, where $E_{sp}$
is the single-particle energy inside the QD, $\gamma=d_1/d_2$ is
the lever arm coefficient, and $d_1$ and $d_2$ are the dimensions
of the structure (fig.~\ref{fig1}). For example, $\gamma\sim 1/6$
in ref.~\cite{QDVoltage3}. For convenience, we assume that the
Fermi energy of the metal contact is zero and loading of the first
hole to the QD occurs at zero bias. Then, the first charged states
of the QD ($n_h=1$ and $2$) have the energies:

\begin{eqnarray}
E_1(V_g) = \gamma |e| V_g+E_{b}^{n_h=1}, \nonumber \\  E_2(V_g)=
2\gamma |e| V_g+U^{dir}_{ss}.  \label{E12a}
\end{eqnarray}
The energies of the next charged states ($n_h=3$ and $4$) are

\begin{eqnarray}
E_3(V_g) = \hbar\omega_0+3\gamma |e| V_g
+U_{coul,3}+E_{b}^{n_h=3}, \nonumber \\  E_4(V_g)=
2\hbar\omega_0+4\gamma |e| V_g +U_{coul,4}+E_{b}^{n_h=4}.
\label{E12b}
\end{eqnarray}
The Coulomb energies $U_{Coul,n_h}$ in the above equations should
be calculated for the ground-state configurations shown in
fig.~\ref{fig5}a. Figure \ref{fig11} shows the calculated energies
of the first changed states related to the s- and p-shells for
non-magnetic QDs. As was realized in several experiments, the
ground state of the system changes with voltage: a QD sequentially
traps $1,2,3,...$ particles \cite{QDVoltage1,QDVoltage3}. Then,
the quantity $D(V_g)$ and the capacitance demonstrate peaks at the
voltages of the ground-state transitions. In real QD systems these
peaks are broadened due to the nonzero size dispersion in a QD
ensemble. The effects of the ferromagnetic interaction are clearly
seen in the calculated capacitance spectra (fig.~\ref{fig12}). The
spacing between the p-orbital peaks $n_h=4$ and $5$ becomes
strongly increased. It comes from the strong ferromagnetic
coupling in the regime of Hund's rule for the state $n_h=4$. The
spacing between the two s-orbital peaks is also increased due to
the MP effect. At the same time, the voltage interval between the
s- and p-related structures becomes reduced. This is again due to
the exchange interaction.

According to fig.~\ref{fig12}, the magnetic and non-magnetic QDs
with the parameters chosen in this paper show the same order of
peaks in the capacitance spectra. With increasing Mn-hole
interaction, the situation can change and the sequence of peaks
can became different for the magnetic and non-magnetic systems.
The reason is that, in magnetic QDs, more carriers can be trapped
to achieve the minimum energy.

In addition to the characteristic behavior of the inter-peak
spacings, the MP effect in the capacitance spectra can be
recognized by varying temperature or by applying an external
magnetic field. With temperature, the peaks related to the most
bound MP states ($n_h=1,3,4,5$) will have the strongest
temperature dispersion. As for the magnetic fields, one
possibility is to use the in-plane field and to suppress the
Mn-hole interaction. Then, the characteristic magnetic dispersion
of the peak positions will reveal the MP binding energy. The
behavior of the peaks in a perpendicular magnetic field can also
be revealing as was shown in other publications \cite{MPBulk}.

\section{Discussion}

The model with exchange interaction of the type
$\hat{j}_{z}\hat{S}_{z,i}\delta(R_h-R_{i})$ assumes that the
magnetic impurity does not create any spin-independent potential.
This model is widely applied for II-VI semiconductors where
individual impurities do not form bound acceptor states. This is
in contrast to the GaAs system.  In the GaAs crystal, a magnetic
Mn impurity forms a deep acceptor state (about $110~meV$ above the
top of the valence band) and therefore the model of impurity in
bulk GaAs  should incorporate the effect of the spin-independent
attracting potential. Treating the typical $III-V$ QDs realized in
the InGaAs system we should take into account two factors: (1) the
possible lower binding energy in the $InGaAs$ system for the Mn
acceptor state \cite{BruceW} and (2) the spatial confinement in a
QD. To understand the importance of the acceptor potential, we
should compare the localization length of the Mn-acceptor in bulk
with the size of the QD.  If the QD dimension is smaller than the
Mn-acceptor size, the QD can be treated without the acceptor
potential. Simultaneously, the spin-dependent exchange interaction
should remain in the model since it leads to the formation of the
MP state. In the opposite limit of a weak QD confinement, the QD
potential can be treated as a perturbation; this case was recently
analyzed in ref.~\cite{Govorov1}. To summarize, if the electronic
size of a QD becomes smaller than the dimension of the acceptor
states of the Mn impurities inside a QD, the simple model of
contact exchange interaction becomes applicable. This suggests
that, under certain conditions, InGaAs QDs can also be treated
with the simple model used in this paper.

To conclude, we have calculated the MP energies and associated
capacitance spectra of QDs with a few holes in the presence of the
Mn-hole exchange interaction. The system studied in this paper
exhibits several features coming from the joint action of the
Mn-hole exchange coupling, quantum confinement, and Coulomb
interaction.

The author would like to thank Bruce McCombe, Leigh Smith, and
Pierre Petroff for motivating discussions. This work was supported
by Ohio University and The Alexander von Humboldt Foundation.


\begin{thebibliography}{99}
\bibitem{Mn-electron} We should note that this is in contrast to the Mn-electron
exchange interaction. The latter is typically isotropic and is
described with the operator $\alpha(\hat{{\bf S}}*\hat{{\bf
s}}_e)$, where $\hat{{\bf s}}_e$ is the spin of an electron in the
conduction band.
\end{thebibliography}
\end{document}